\documentclass[prb,aps,twocolumn,floats,superscriptaddress,longbibliography]{revtex4-2}
\usepackage{soul}
\usepackage{xcolor}
\usepackage{multirow}
\usepackage{manfnt}
\usepackage{amsmath}
\usepackage{graphicx}
\usepackage{amssymb}
\usepackage{epsfig}
\usepackage{simplewick}
\usepackage[colorlinks = true]{hyperref}
\begin{document}
	\newlength{\upit}\upit=0.1truein
\newcommand{\cube}{\mbox{\mancube}}
	\newcommand{\raiser}[1]{\raisebox{\upit}[0cm][0cm]{#1}}
	\newcommand{\ltappr}{{{\lower4pt\hbox{$<$} } \atop \widetilde{ \ \ \ }}}
	\newlength{\bxwidth}\bxwidth=1.5 truein
	\newcommand\frm[1]{\epsfig{file=#1,width=\bxwidth}}
	\newcommand{\cg}{{\cal G}}
	\newcommand{\dif}[2]{\frac{\delta #1}{\delta #2}}
	\newcommand{\ddif}[2]{\frac{\partial #1}{\partial #2}}
	\newcommand{\Dif}[2]{\frac{d #1}{d #2}}
	\newcommand{\str}{\hbox{Str}}
	\newcommand{\Str}{\underline{\hbox{Str}}}
	\newcommand{\tr}{{\hbox{Tr}}}
	\newcommand{\Tr}{\underline{\hbox{Tr}}}
	\newcommand{\dg}{^{\dagger }}
	\newcommand{\vk}{\mathbf k}
	\newcommand{\vq}{{\vec{q}}}
	\newcommand{\vp}{\bf{p}}
	\newcommand{\al}{\alpha}
	\newcommand{\BigL}{\biggl }
	\newcommand{\BigR}{\biggr }
	\newcommand{\gtappr}{{{\lower4pt\hbox{$>$} } \atop \widetilde{ \ \ \ }}}
	\newcommand{\si}{\sigma}
	\newcommand{\rarrow}{\rightarrow}
	\newcommand{\up}{\uparrow}
	\newcommand{\dw}{\downarrow}
	\def\fig#1#2{\includegraphics[height=#1]{#2}}
	\def\figx#1#2{\includegraphics[width=#1]{#2}}

	\newcommand{\bk}{{\bf{k}}}
	\newcommand{\bx}{{\bf{x}}}
\newcommand{\ba}{{\bf{a}}}
	\newcommand{\pmat}[1]{\begin{pmatrix}#1\end{pmatrix}}
	\newcommand{\ua}{\uparrow}
	\newcommand{\da}{\downarrow}
	
	\newcommand{\be}{\begin{equation}}
		\newcommand{\ben}{\begin{equation*}}
			\newcommand{\ee}{\end{equation}}
		\newcommand{\een}{\end{equation*}}
	\newcommand{\parr}{\parallel}
	\newcommand{\bmx}{\begin{array}}
		\newcommand{\emx}{\end{array}}
	\newcommand{\bean}{\begin{eqnarray*}}
		\newcommand{\eean}{\end{eqnarray*}}
	\newcommand{\dn}{^{\vphantom{\dagger}}}
	\newcommand{\nn}{\vphantom{-}}
	\newcommand{\lr}{\leftrightarrow}
	\newcommand{\ra}{\rightarrow}
	\newcommand{\la}{\leftarrow}
	\newcommand{\bb}[1]{\mathbb{#1}}
	\newcommand{\qqquad}{\qquad\qquad\qquad}
	\newcommand{\eps}{\epsilon}
	\newcommand{\sgn}[1]{{\rm sign}{#1}}
	\newcommand{\pref}[1]{(\ref{#1})}
	\newcommand{\tilda}[2]{\intopi{d{#1}}\Big(g_{#1}^A(\eps_p){#2}\Big)}
	\newcommand{\intpi}[1]{\int_{-\pi}^{+\pi}{#1}}
	\newcommand{\im}[1]{{\rm Im}\left[ #1 \right]}
	\newcommand{\trr}[1]{{\rm Tr}\Big[ #1 \Big]}
	
	\newcommand{\abs}[1]{\left\vert #1 \right\vert}
	\newcommand{\bra}[1]{\left\langle #1 \right\vert}
	\newcommand{\ket}[1]{\left\vert #1\right\rangle}
	\newcommand{\braket}[1]{\left\langle #1\right\rangle}
	\newcommand{\mat}[1]{\left(\bmx{cc}#1\emx\right)}
	\newcommand{\matc}[2]{\left(\bmx{#1}#2\emx\right)}
	\newcommand{\matn}[1]{\bmx{cc}#1\emx}
	\newcommand{\matl}[1]{\bmx{ll}#1\emx}
	\newcommand{\sepline}{\begin{center}\rule{8cm}{.5pt}\end{center}}
	\newcommand{\hi}{\noindent\currfilebase.pdf \hspace{.2cm}-\hspace{.2cm} \today}
	\newcommand{\hiy}{\noindent Yashar \currfilebase.pdf \hspace{.2cm}-\hspace{.2cm} \today}
	
	\setlength{\parindent}{0.5cm}
	\newcommand{\indentoff}{\setlength{\parindent}{0cm}}
	\newcommand{\EJK}[1]{{\color{olive} EJK: #1}}
	\newcommand{\imEJK}[1]{{\color{olive}\bf \Large \ddag}\marginpar{\scriptsize \color{olive}\bf  \ddag #1}}
	\newcommand{\SEC}[1]{{\color{blue} \textit{#1}}}
	\newcommand{\red}[1]{{\color{red}#1}}
	\newcommand{\blue}[1]{{\color{blue}#1}}
	\newcommand\relbd{\mathrel{{\bf\smash{{\phantom- \above1pt \phantom-
	}}}}}
	\newcommand\ltdash{\raise-0.7pt\hbox{$\scriptscriptstyle |$}}
	\def\fig#1#2{\includegraphics[height=#1]{#2}}
	\def\figx#1#2{\includegraphics[width=#1]{#2}}
	\newlength{\figwidth}
	\figwidth=10cm
	\newlength{\shift}
	\shift=-0.2cm
	\newcommand{\fg}[3]
	{
		\begin{figure}[ht]
			
			\vspace*{-0cm}
			\[
			\includegraphics[width=\figwidth]{#1}
			\]
			\vskip -0.2cm

			\caption{\label{#2}
				\small#3
			}
	\end{figure}}
	\newcommand{\fgb}[3]
	{
		\begin{figure}[b]
			\vskip 0.0cm
			\begin{equation}\label{}
				\includegraphics[width=\figwidth]{#1}
			\end{equation}
			\vskip -0.2cm
			\caption{\label{#2}
				\small#3
			}
	\end{figure}}
	\graphicspath{{Figures/CPTVfigs/}}
	
	\title{Microscopic theory of pair density waves in spin-orbit coupled Kondo lattice
	}
	\author{Aaditya Panigrahi}
	\affiliation{
		Center for Materials Theory, Department of Physics and Astronomy,
		Rutgers University, 136 Frelinghuysen Rd., Piscataway, NJ 08854-8019, USA}
  \author{Alexei Tsvelik}
	\affiliation{Division of Condensed Matter Physics and Materials Science, Brookhaven National Laboratory, Upton, NY 11973-5000, USA}
\author{Piers Coleman}
	\affiliation{
		Center for Materials Theory, Department of Physics and Astronomy,
		Rutgers University, 136 Frelinghuysen Rd., Piscataway, NJ 08854-8019, USA}
	\affiliation{Department of Physics, Royal Holloway, University
		of London, Egham, Surrey TW20 0EX, UK.}
	\date{\today}
	
	\pacs{PACS TODO}
	\begin{abstract}
		{ 
			\begin{center}
            \begin{minipage}{0.8\textwidth}
                {
            We demonstrate that the discommensuration
            between the Fermi surfaces of a conduction sea and an underlying spin liquid provides a natural mechanism for the spontaneous formation of pair
            density waves.  Using a recent formulation of the Kondo lattice model which incorporates a Yao Lee spin liquid proposed by the authors,  we demonstrate that doping away from half-filling induces finite-momentum electron-Majorana pair condensation, resulting in amplitude-modulated PDWs. Our approach  provides a precise, analytically tractable pathway for understanding the spontaneous formation of PDWs in higher dimensions and offers a natural mechanism for PDW formation in the absence of Zeeman splitting.}
                \end{minipage}
			\end{center}
		}
	\end{abstract}
	\maketitle
    \newpage

{\it Introduction.} Spatially modulated pair condensates, or pair density waves\cite{agterberg_physics_2020} (PDWs) were originally envisioned by Fulde, Ferrell\cite{fulde_superconductivity_1964}, Larkin and Ovchinikov\cite{larkin_nonuniform_1964}(FFLO), as a high-field superconducting response to the Zeeman splitting of the Fermi surface.  The recent observation of PDWs via scanning tunneling microscopy\cite{randeria_scanning_2016} measurements in  cuprate\cite{hucker_stripe_2011}, iron-based \cite{zhao_smectic_2023,liu_pair_2023} and heavy fermion superconductors\cite{gu_detection_2023} has sparked a resurgence of interest in this phenomenon.  Crucially, in each of these unconventional superconductors the PDW develops without a magnetic field. 
PDWs naturally form as a  response of  a uniform pair condensate to a charge or spin density wave(CDW)\cite{agterberg_physics_2020,fradkin_stripe_2015}.  However, the recent PDW observation in the novel triplet superconductor, UTe$_2$\cite{gu_detection_2023} found that the PDW survives the melting of CDW order\cite{aishwarya_melting_2024}, suggesting that in this case, the PDW is a primary order parameter, developing  without the aid of a charge or spin density wave.  These observations motivate us to seek a natural mechanism for the spontaneous development of pair density wave order.




The condensation of a PDW requires a pair susceptibility that peaks at a finite momentum. This does not occur naturally in a BCS superconductor\cite{BCS_Theory}, where the logarithmic  divergence of the uniform pair susceptibility is cut off by the momentum, or more precisely by $v_F Q$, where $v_F$ is the Fermi velocity and $Q$ the wavevector of the PDW. In a BCS superconductor, a Zeeman field polarizes the Fermi surfaces, cutting off the zero-momentum instability and favoring a finite-momentum peak in the pair susceptibility, leading to an FFLO-like pair density waves\cite{fulde_superconductivity_1964,larkin_nonuniform_1964}.
While experiment suggests that pair density waves can spontaneously develop in unconventional superconductors even in the absence of a magnetic field, there is  no current consensus on the underlying mechanism. 

In this paper we explore PDW formation in the context of the Kondo lattice model for heavy fermions. The current authors (CPT) have recently proposed a three dimensional, solvable model in which the Kondo screening of a $\mathbb{Z}_2$ spin liquid induces superconductivity in the surrounding conduction sea\cite{coleman_solvable_2022,tsvelik_order_2022,seifert_fractionalized_2018,choi_topological_2018-1,de_carvalho_odd-frequency_2021}. The underlying spin liquid is a three dimensional generalization of the spin-orbital derivative of the well-known $\mathbb{Z}_2$ Kitaev spin liquid, known as the Yao-Lee spin liquid\cite{yao_fermionic_2011}. At half-filling and in the weak coupling limit, Kondo coupling between the spin liquid and the conduction electrons induces odd-frequency triplet superconductivity.

  In this letter, we demonstrate that our model\cite{coleman_solvable_2022} provides a natural mechanism for the emergence of incommensurate pair-density waves when the system is doped away from half-filling. In particular, its weak coupling description offers a long-sought pathway to understanding the spontaneous formation of pair-density waves beyond one dimension \cite{agterberg_physics_2020} in the absence of a field.

    \begin{figure}[h]
        \includegraphics[width=1.\linewidth]{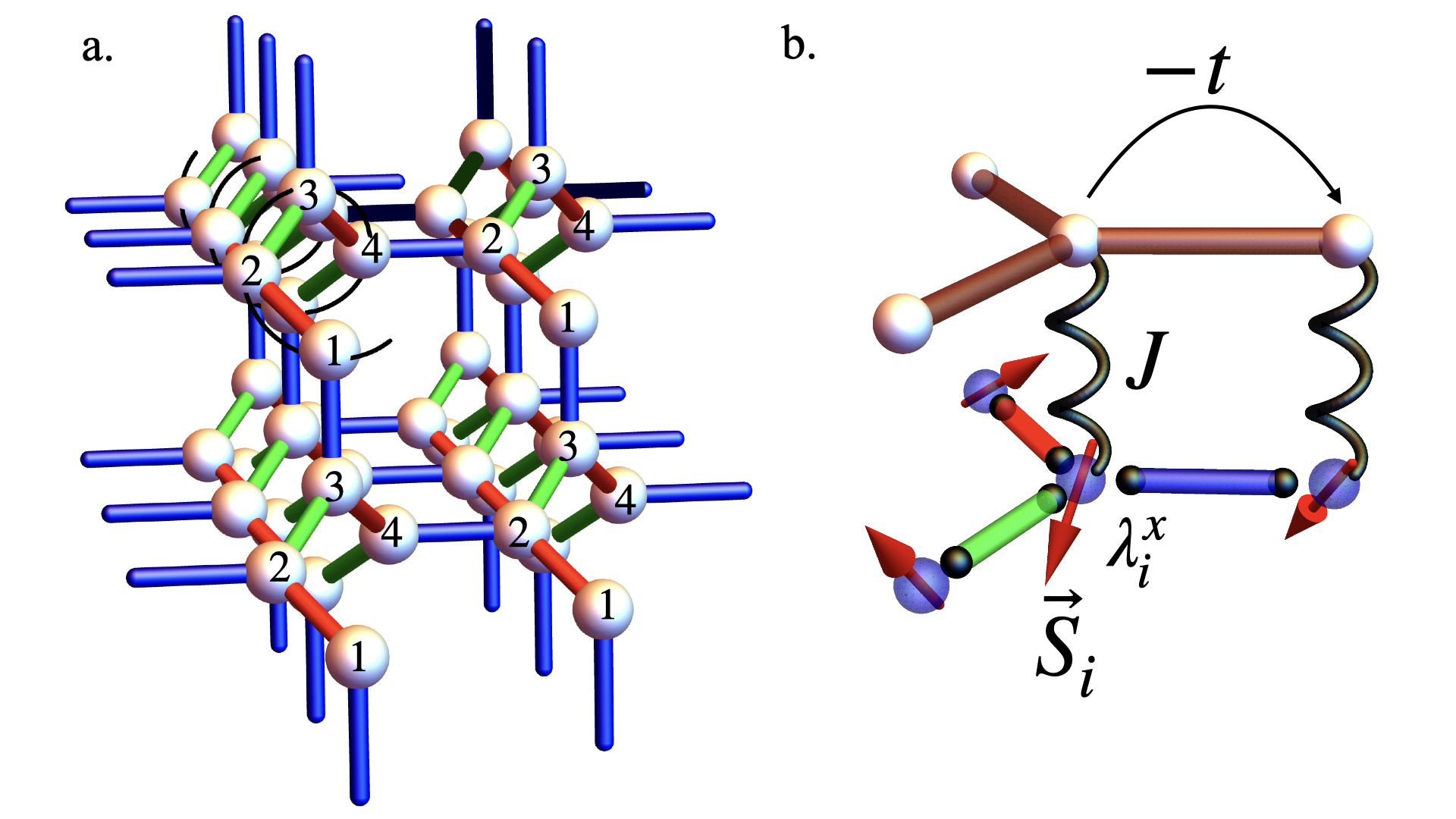}
        \caption{  (a) The hyper-octagon lattice, which features a body-centered cubic (BCC) structure formed by interpenetrating square and octagonal spirals, with a four-atom unit cell \cite{hermanns_quantum_2014}. The frustrated orbital interactions are denoted by color-coded red, blue and green bonds. (b) In the CPT model each site is trivalent, hosting conduction electrons, shown here linked by golden bonds, localized spins and orbitals, linked by color-coded orbital interactions. The frustrated orbital interactions induce a Majorana fractionalization of spins, which are then Kondo-coupled to the conduction electrons on the same lattice. }
        \label{fig:Hyperoctagon}
    \end{figure}

 At half-filling, the model exhibits nesting between electron, hole, and Majorana Fermi surfaces, leading to a logarithmic instability into an electron-Majorana pair condensate under infinitesimal Kondo coupling. 
Our key result  is that doping the system away from half-filling, through a shift in the chemical potential, modifies the electron and hole Fermi surfaces while leaving the Majorana Fermi surface unchanged. This imbalance induces a finite-momentum FFLO-like electron-Majorana condensation, giving rise to amplitude-modulated pair-density waves.

A key advantage of our model over general parton theories is that it does not rely on an approximate Gutzwiller projection to enforce the single-occupancy constraint by mean-field methods. Instead, in the Yao-Lee spin liquid, this constraint is inherently encoded in the Majorana representation, offering a more precise understanding of how the chemical potential influences the order parameter.

 {\it Model.} The CPT model\cite{coleman_solvable_2022}, defined on the hyperoctagon lattice (Fig. \ref{fig:Hyperoctagon}), Kondo-couples a band of conduction electrons to the spins of a Yao-Lee spin liquid. Each site possesses three degrees of freedom: electrons, localized spins, and localized orbitals, leading to a Hamiltonian with three components, $H_{CPT}=H_c+H_{YL}+H_K,
			\label{CPTHam} 
        $
where $H_c$ describes the nearest-neighbor hopping of the conduction electrons, $H_{YL}$ captures the Yao-Lee spin-spin interaction, and $H_K$ couples the conduction sea to the Yao-Lee spin liquid through the Kondo interaction,
        \begin{align}
			H_{c}=&-t\sum_{\braket{i,j}} ( c\dg_{i\si}c_{j\si}+{\rm H.c.})-\mu\sum_{i}c\dg_{i\si}c_{i\si},\nonumber
			\\H_{YL}=&K/2\sum_{\braket{i,j}}\lambda^{\al_{ij}}_i\lambda^{\al_{ij}}_j
			(\vec{S}_i\cdot\vec{S}_j),\label{YLH}
			\\H_{K}=&J\sum_{i}(c\dg_{i}\vec{\si}c_{i})\cdot \vec{S}_{i}.
			\label{CPTComp} \nonumber
        \end{align} 
The Yao-Lee term involves an anisotropic nearest-neighbor Ising interaction between the $\alpha_{ij} = x, y, z$ components of the orbitals $\lambda_j$, which is decorated by a Heisenberg interaction between the spins $\vec{S}_j$. 
 The  orbital frustration in $H_{YL}$ induces large quantum fluctuations, leading to a fractionalization of the spins and orbitals into Majorana fermions as follows,  $\vec{\lambda}_j = -i \vec{b}_j \times \vec{b}_j$ and  $\vec{S}_j = -\frac{i}{2} \vec{\chi}_j \times \vec{\chi}_j$, where $b$ and $\chi$ are Majorana fermions. At low energies, the orbital degrees of freedom decouple as static $\mathbb{Z}_2$ gauge fields  which  freeze into a flux-free configuration at low temperatures, allowing  the low energy physics of the Yao-Lee spin liquid to  be described by a single-band Hamiltonian
  \begin{equation}\label{MajoranaHam}
    H_{YL}=\sum_{{\bf k}\in \cube} \epsilon_{\chi}({\bf k})  \vec{\chi}^{\dg}_{{\bf k}} \cdot \vec{\chi}_{{\bf k}},
\end{equation}
where the  momentum sum takes place over the cubic Majorana Brillouin zone $\cube$ and the 
dispersion $\epsilon_{\chi}({\bf k})=K\tilde{\epsilon}({\bf k})$, where 
  \begin{equation} \label{ProjectedBand2}
 \tilde{\epsilon}({\bf k})=\frac{1}{2 s_{x}s_{y}s_{z}} \left[(c^2_x +c^2_y +c^2_z)-\frac{3}{4}\right],
\end{equation}
where  $c_i=\cos(k_i/2)$ and $s_i=\sin(k_i/2)$.

        \begin{figure}[h]
            \includegraphics[width=1.0\linewidth]{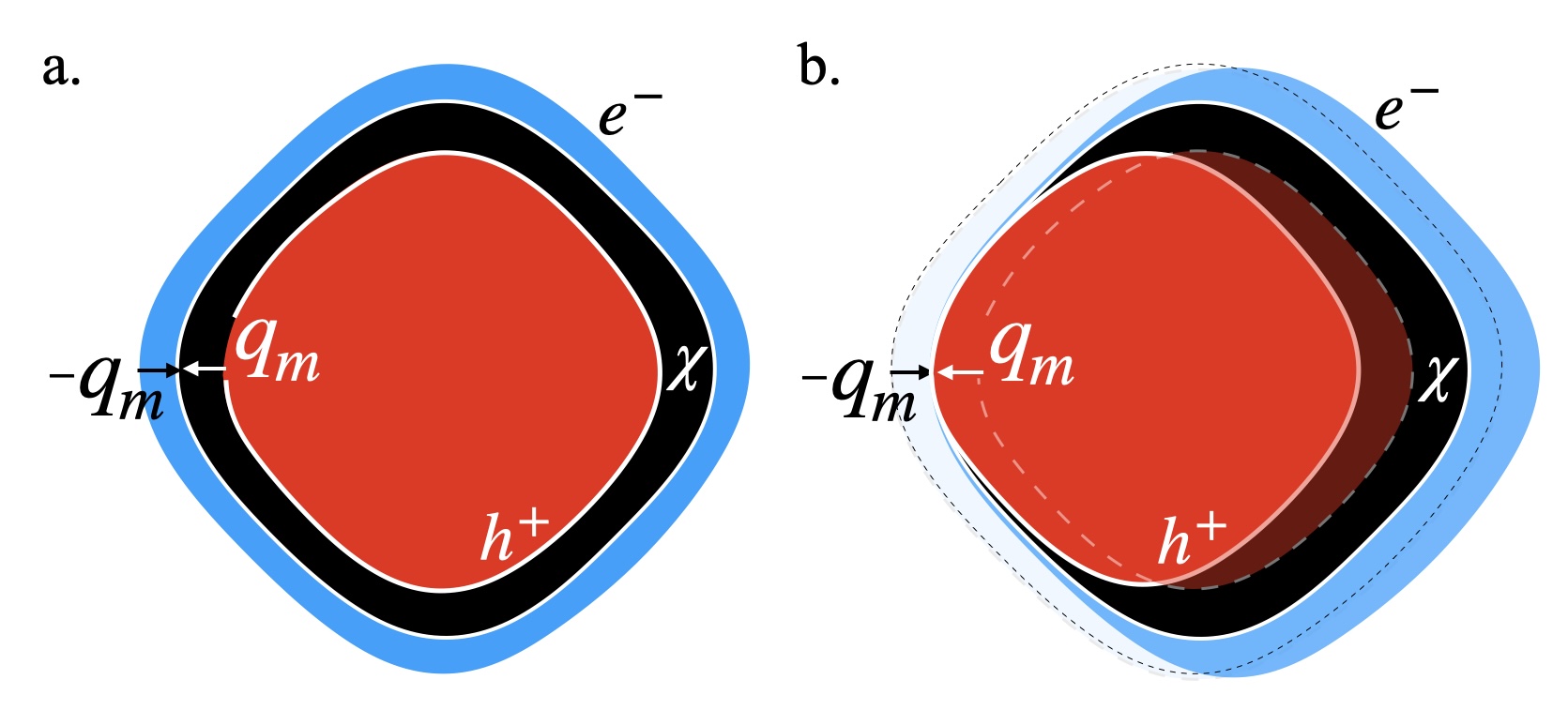}
            \caption{Illustrates the impact of electron doping on the conduction sea within a hyperoctagon lattice. (a)The conduction sea consists of  electron(blue) and hole Fermi surfaces (red), which expand and contract, respectively, as electron doping shifts the system away from half-filling. By contrast, the Majorana Fermi surface (black) is unaffected by doping.   (b) Doping stabilizes the development  of finite-momentum configurations of the electron-Majorana spinor order by shifting  the electron and hole Fermi surfaces shift by $\pm\vec{q}_m$ to nest with the Majorana Fermi surface.
            }
            \label{fig:Fermi_Surface}
        \end{figure}
Similarly, the low energy physics of the conduction band\cite{coleman_solvable_2022} can be described by a  single band $\epsilon_c({\bf k})=-t \tilde{\epsilon}({\bf k})$, with $\tilde{\epsilon}({\bf k})$ taking the form (\ref{ProjectedBand2}). By rewriting the conduction Hamiltonian in terms of the Balian-Werthammer spinor  $\psi_{\bf k}=(c_{{\bf k}}, -i \sigma^2 c^{\dg}_{-{\bf k}})^T$, the conduction sea Hamiltonian becomes,
\begin{equation} \label{BLHcond}
    H_{c}=\sum_{{\bf k}\in \cube}\psi^{\dagger}_{{\bf k}}(\epsilon_{c}({\bf k})\mathbb{I}-\mu  \tau_3)\psi_{{\bf k}}.
\end{equation}
The Majorana fractionalization of the spins enables Kondo interaction to be rewritten in terms of Majorana fermions and conduction electrons as follows
	\begin{equation}\label{KondoInteraction}\small
		H_{K}=
 - \frac{J}{2}\sum_{l}c\dg_{j}
(\vec{\chi }\cdot \vec{ \sigma })^{2}c_{j}.
	\end{equation}
 $H_{K}$ can be decoupled via a Hubbard-Stratonovich transformation in terms of a spinor order parameter field $V_j = -J \langle \vec{\chi}_j \cdot \vec{\sigma} c_j \rangle = (V_{j\uparrow}, V_{j\downarrow})^{T}$, which is solved self-consistently via a compact mean-field solution for the Kondo interaction. 
    \begin{equation}\label{MFTHam}\small 
    H_{int}[j]=\frac{1}{2}\left(\psi^{\dg}_j (\vec{\si}\cdot \vec{\chi}_j) \mathcal{V}_j+\mathcal{V}^{\dg}_{j}(\vec{\si}\cdot\vec{\chi}_j)\psi_j \right)+\frac{\mathcal{V}^{\dg}_j \mathcal{V}_j}{J}, 
\end{equation}
where $\psi_{\vec{k}} = (c_{\vec{k}}, -i \sigma^2 c^{\dagger}_{-\vec{k}})^T$ and $\mathcal{V}_j = (V_j, -i \sigma^2 V^*_j)^T$.
 In previous work \cite{coleman_solvable_2022}, we showed that at half-filling, the model exhibits a logarithmic singularity in the electron-Majorana pair susceptibility, leading to odd-frequency triplet superconductivity.  

{\it Pair density waves.}
We now show that doping provides a natural mechanism for the spontaneous formation of a pair density wave (PDW) { with a modulated gap function},  in which the chemical potential plays a role analogous to a Zeeman field in conventional superconductors. The introduction of chemical potential splits the electron and hole Fermi surfaces while leaving the Majorana Fermi surface unaffected (Fig \ref{fig:Fermi_Surface}a). 
This then causes the spinor order parameter $\mathcal{V}_{\vec{q}}$ to acquire a finite momentum $\vec{q}$ (Fig. \ref{fig:Fermi_Surface}b), resulting in the formation of a PDW. The Fourier transform of the Kondo interaction with finite momentum order parameter is given by,
        \begin{equation}\small
    H_{int}=\sum_{\vec{k}\in \cube}\left(\psi^{\dg}_{\vec{k}+\vec{q}}(\vec{\si}\cdot \vec{\chi}_{\vec{k}})\mathcal{V}_{\vec{q}}+h.c.\right)+\mathcal{N}\frac{\mathcal{V}_{\vec{q}}^{\dg}\mathcal{V}_{\vec{q}}}{J}.
        \end{equation}
        
\begin{figure}[h]
            \includegraphics[width=1.\linewidth]{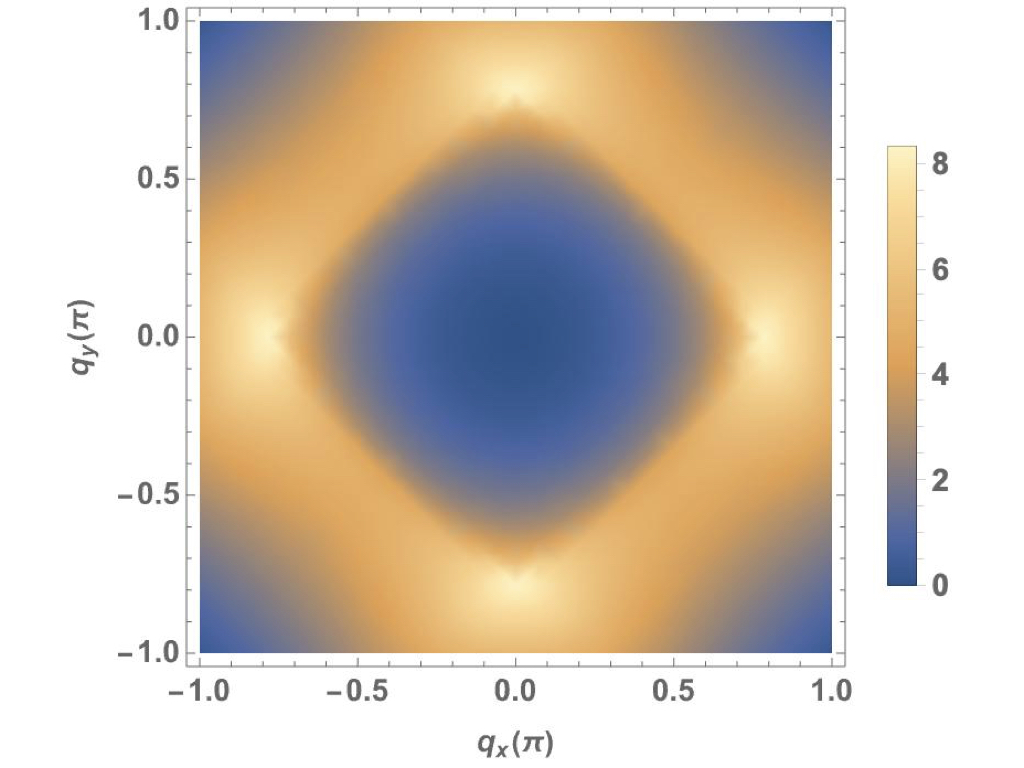}
            \caption{ Depicts the finite momentum susceptibility $\vec{q} = \vec{Q} - \vec{P}$ for the formation of electron-Majorana spinor order $\mathcal{V}_{\vec{Q}}$ along the $q_z = 0$ plane. The susceptibility reaches its maximum at $\vec{q}_i = q_m(\pm 1, 0, 0), q_m(0, \pm 1, 0)$, where $q_m = 0.7738\pi \mu/t$, illustrating the four-fold degeneracy. Due to cubic symmetry, the susceptibility is also maximized at $\vec{q}_i \in \{q_m(\pm 1, 0, 0), q_m(0, \pm 1, 0), q_m(0, 0, \pm 1)\}$.
 }
            \label{fig:Susceptibility1}
        \end{figure}

\begin{widetext}
Finite-momentum spinor order arises when the electron-Majorana pair susceptibility (Fig. \ref{fig:Susceptibility1}) 
        \begin{equation}\label{Susceptibility1}\small
    \frac{\partial^2 F_e}{\partial V^2}=\chi_{P}(\vec{q})=-{2T}\sum_{i\omega_n,k\in \cube}\Tr\left[G_{\psi}(\vec{k}+\vec{q},i\omega_n)\sigma^a \mathcal{Z}\mathcal G_{\chi^a}(\vec{k},i \omega_n)\mathcal{Z}^{\dagger}\sigma^a\right],
        \end{equation}
where $F_e$ is the electronic part of the free-energy, develops a maximum at a finite momentum $\vec{q}$. Here $G_{\psi}(\vec{k}+\vec{q},i\omega_n)$ and $G_{\chi}(\vec{k},i\omega_n)$ are the
conduction and Majorana Green's functions respectively, with $\mathcal{Z}=\frac{\mathcal{V}}{\sqrt{2}V}$  representing the normalized orientation of the spinor order (${\mathcal Z}\dg \mathcal{Z}=1$). Further simplifications, outlined in the supplementary material, allow us to  rewrite the susceptibility for the hyperoctagon lattice in the following closed form,
\begin{equation}\label{HyperoctagonSusceptibilty2}
\small \chi_P (\vec{q})=\frac{3\rho(0)}{(K+t)}\ln\left(\frac{(K+t) D}{\mu}\right)-\frac{3\rho(0)}{4(K+t)}\int \frac{d\Omega}{4\pi}\left[\ln\left(\left\vert 1-\frac{\vec{v}_F(\theta,\phi)\cdot \vec{q}}{\mu}\right\vert\right)+\ln\left(\left\vert \frac{K}{t}+ \frac{\vec{v}_F(\theta,\phi)\cdot \vec{q}}{\mu}\right\vert \right)\right].
\end{equation}
\end{widetext}

Here, $\rho(0)$ is the density of states at the Fermi surface, and $\vec{v}_F=\vec{\nabla}_{\bf k} \epsilon_{c}({\bf k})$ is the Fermi velocity of the conduction sea, which, for the hyperoctagon lattice, depends on the orientation $(\theta,\phi)$. The logarithmic instability of the electron-Majorana susceptibility in equation (\ref{HyperoctagonSusceptibilty2}) is cut off by the chemical potential, which splits the nesting between the conduction sea and the Majorana Fermi surfaces. This results in the susceptibility maxima developing at a finite momenta $\vec{q}$ (Fig. \ref{fig:Susceptibility1}). 

To evaluate the susceptibility in angular coordinates, we rewrite the Fermi-velocity of the conduction sea in the momentum basis,
\begin{equation}\label{FermiVelocity2}
    \vec{v}_F=\frac{1}{2 s_x s_y s_z}\left(\frac{c_x^3}{s_x},\frac{c_y^3}{s_y},\frac{c_z^3}{s_z}\right),
\end{equation}
where, $c_i=\cos(k_i/2)$ and $s_i=\sin(k_i/2)$.
We then parameterize the $c_i$ of the Fermi surface in  spherical coordinates 
\begin{equation}\label{angCos}\small
    (c_x,c_y,c_z)=(r\cos(\phi)\sin(\theta),r\sin(\phi)\sin(\theta),r\cos(\theta)),
\end{equation}
in terms of angular variables $\theta$ and $\phi$ and  radius $r=\sqrt{3}/2$.  
Carrying out the angular integral in equation (\ref{HyperoctagonSusceptibilty2}), we numerically find that the susceptibility is maximized (Fig. \ref{fig:Susceptibility1}) for:
\begin{equation}
\vec{q}_i \in {q_m(\pm 1,0,0), q_m(0,\pm 1,0), q_m(0,0,\pm 1)},
\label{maxima_vectors}
\end{equation}
along the principal axes where the magnitude of the wavenumber at the susceptibility maxima $q_m = 0.77\pi \mu$ (Fig. \ref{fig:Susceptibility1}) is proportional to the chemical potential.

\begin{figure}[h]
            \includegraphics[width=1.\linewidth]{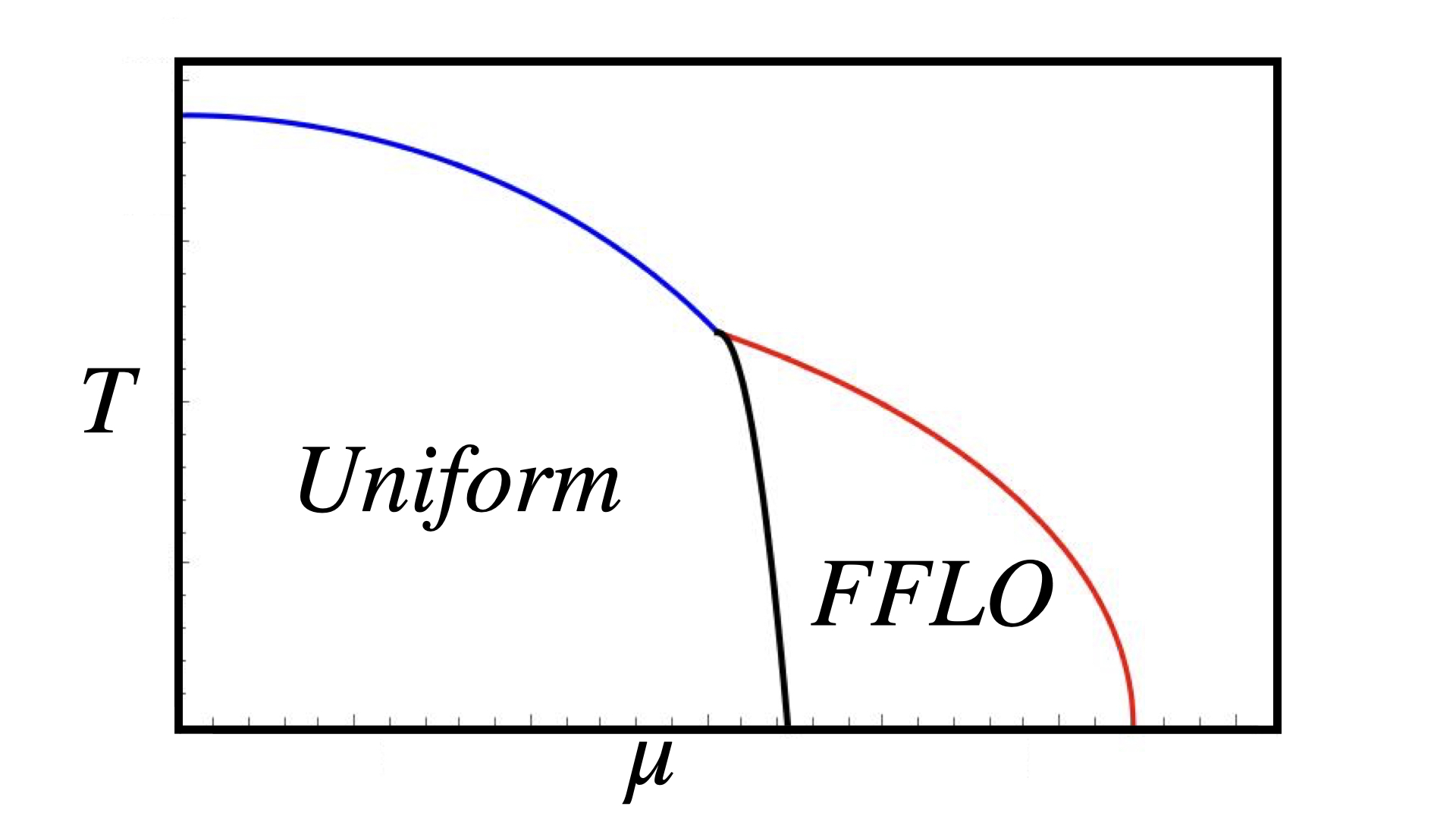}
            \caption{Depicts the phase diagram of the CPT model as a function of  temperature $T$, and chemical potential $\mu$ for a finite Kondo coupling $J$. Second-order transitions separate the normal state from the uniform superconducting (blue) and incommensurate FFLO-like PDW phases(red). The first-order boundary (black) between the two ordered phases appears where the critical temperatures coincide. This black line terminates at a critical chemical potential $\mu_c$, marking a first-order transition between the two phases at zero temperature.
 }
            \label{fig:PhaseDiagram}
        \end{figure}

At zero temperature, the pairing susceptibility $\chi_P(\vec q_m) $ peaks at finite momentum of magnitude $|\vec{q}_m| \propto \mu$, favoring finite-momentum order. At the superconducting transition temperature $T_c$, staggered order develops provided $\chi(T,\vec{q})$ is maximal at a nonzero momentum $\vec{q}_m$, implying convexity of the pair susceptibility, $b(T,\mu) = d^2 \chi(T,q)/dq^2\bigr\vert _{q=0} >0$ at  $\vec{q} = 0$. Detailed calculation shows that $b(T,\mu) > 0$ is positive provided the chemical potential exceeds  $\mu_c=3.82T$ (see Supplementary Material). Beyond this point,  the system undergoes a second-order transition from the normal state into an incommensurate PDW upon cooling. In {the absence of fluctuation,} mean-field theory {predicts that} the uniform state at $\mu<\mu_c$ and the PDW at $\mu>\mu_c$ are separated by a first-order transition (Fig.\ref{fig:PhaseDiagram}). 
{The role of long-wavelength amplitude fluctuations in shaping this phase boundary is left for a future work.}
 

In the PDW phase, the cubic symmetry of our model leads to six degenerate wave vectors along which the spinor order prefers to modulate. Such degeneracy is often lifted, as in the case of FFLO\cite{larkin_nonuniform_1964} states, where the order parameter favors amplitude modulation due to attractive interactions between $\mathcal{V}_{\vec{q}}$ and $\mathcal{V}_{-\vec{q}}$. This results in ordering of the form $\mathcal{V}_{\vec{q}}e^{i \vec q \cdot \vec x} + \mathcal{V}_{-\vec{q}}e^{-i \vec q \cdot \vec x}$, or other superpositions, depending on the level of doping.

\vskip 0.1in

{\it Landau Theory.} 

\begin{widetext}

    In the incommensurate phase there will in general be up to six modulation vectors $q_i$ ($i=1,6$).  The Landau theory describing the energetics of this state depends on the six spinors ${\cal V}_{q_i}$ and has the general form 
     \begin{align}
\small \label{LandauGinzburgZ}
    \mathcal{F}=&\sum_{i}\Big[\tau ({\cal V}^{\dg}_{q_i}{\cal V}_{q_i}) 
    +\frac{\beta_1}{2} ({\cal V}^{\dg}_{q_i}{\cal V}_{q_i})^2
     \Big] +\sum_{ i\neq j}\frac{{\beta_2}}{2} ({\cal V}^{\dg}_{q_i}{\cal V}_{q_i}) ({\cal V}^{\dg}_{q_j}{\cal V}_{q_j}) +\frac{{\beta_3}}{2} ({\cal V}^{\dg}_{q_i}{\cal V}_{q_j})({\cal V}^{\dg}_{q_j}{\cal V}_{q_i})  + \frac{\beta_4}{2} ({\cal V}^{\dg}_{q_i} {\cal V}_{q_j})( {\cal} {\cal V}^{\dg}_{-q_i} {\cal V}_{-q_j}).
\end{align}
 Minimizing (\ref{LandauGinzburgZ}) with respect to ${\cal V}_{q_i}$ determines the stable configurations of the PDW. 

\end{widetext}

The quadratic degeneracy amongst the order parameters $\mathcal{V}_{\vec{q}_i}$ is broken by the quartic coefficients $\beta_i$, allowing for solutions in which multiple $V_{q_i}$ acquire an expectation value. 
 Let us consider the simplest PDW with  unidirectional modulation in which the ${\cal V}_{ q_i}= V {\cal Z}_{q_i}$ ($i=1,2$).   In general, we have to allow for the possibility that the two spinors have a different orientation, so that ${\cal Z}\dg _{q_i}{\cal Z}_{-q_i}= \cos^2\phi/2$ is not equal to unity.  The modulated phase becomes
stable relative to the unmodulated case provided $\beta_2 + \beta_3 \cos^2 \phi/2< \beta_1$, moreover, if $\beta_3>0$, then $
\phi\neq 0$ and the two spinors in the PDW are no longer aligned.  We will shortly see that even when $\phi = \pi$
and the spinors are orthogonal, the electron self energy is still modulated.   More complex intertwined order is possible, giving rise in general to a multi-wavevector  PDW 
\begin{eqnarray}
  {\cal V}(\vec{x}) = V_0\sum_i{\cal Z}_i e^{i \vec q_i\vec x}, ~~{\cal Z}_i^+{\cal Z}_i =1.  
  \label{spinororder}
\end{eqnarray}
Unlike  standard models\cite{fulde_superconductivity_1964,larkin_nonuniform_1964}, this triplet PDW is described in terms of the six independent spinors ${\cal Z}_i$ enabling a rich variety of modulated orders.

{\it Modulated pairing and intertwined order. }  
{
The  modulation in the electronic response that arises from the modulated spinor ordering is encoded in the electronic self-energy, obtained by integrating out the Majorana fermions,
\begin{equation}\Sigma(x,x',\omega) = \sigma_a\mathcal{V}(x) G_{\chi}(x-x',\omega) \mathcal{V}\dg(x')\sigma_a.
\label{self-energy}
\end{equation}
The local component of this self-energy, obtained by setting $x=x'$, can be utilized to study the modulated order resulting from the PDW(\ref{spinororder}). 

To illustrate this point, we consider the  simplest PDW with unidirectional modulation which involves two condensed spinors so that  $\mathcal{Z}_i,\mathcal{Z}_{-i}\neq 0$. Depending on the coefficient $\beta_3$ of the Landau theory,  $\mathcal Z_i$ and $\mathcal Z_{-i}$ may not be equal.  The various intertwined components of the modulated order can be understood by decomposing the self-energy,
\begin{equation}  
    \Sigma(x,\omega) 
    = \Sigma_{\alpha,\beta}(x,\omega)\sigma^{\alpha}\tau^{\beta}, \qquad (\alpha,\beta \in 0,3)
    \label{self-energy}  
\end{equation}  
in terms of the spins and iso-spin operators $\si^{\al}$ and $\tau^{\beta}$ respectively. 

The components of this four-dimensional matrix describe an intertwined order between
the primary, modulated triplet pairing and secondary charge, singlet pair and spin density waves. 
For example, the spacelike  components $\Sigma_{ab}(x,\omega$) ($a,b =1,2,3$) define the primary, modulated triplet order with a d-vector  $[\vec{d}^b]_a(x,\omega)=\Sigma_{ab}(x,\omega) \propto \cos (2 \vec q \cdot \vec x)$

By contrast, the mixed components $\Sigma_{0a}(\vec x,\omega)$ and $\Sigma_{a0}(\vec x,\omega)$ describe secondary charge, singlet pair and spin density wave components. Terms with only a single $\sigma^a$
or $\tau^b$ pick up a minus sign under transposition, which causes these components to be proportional to $\sin (2 \vec q \cdot \vec x)$, so the secondary order is $\pi/2$ out of phase with the primary order (see supplementary materials for details).  Thus  $\Sigma_{03}(\vec x, \omega) \sim \sin (2 \vec q \cdot \vec x) $ describes a staggered scattering potential while $\Delta(\vec x,\omega) = \Sigma_{01}(\vec x,\omega) + i \Sigma_{02}(\vec x,\omega)\sim \sin (2 \vec q \cdot \vec x)  $ describes a modulated s-wave pairing, resulting from the absence of inversion symmetry. Similarly, the  spin-density wave component $\Sigma_{a0}(x,\omega) = \vec M_{a}(\omega)\sin (2 \vec q \cdot \vec x)$ results from the time-reversal symmetry breaking of the spinor order. 
We note that {when the two spinors  $\mathcal Z_{\pm q}$ are aligned, only the triplet and resonant scattering components of the self-energy, i.e., $\Sigma_{\mu\mu}$, remain finite.  The appearance of the same modulation vector in the out-of-phase pair density wave and charge density wave is a unique feature of this PDW mechanism.


\vskip 0.1in
{\it Discussion.}
\vskip 0.1in

In this paper, we have proposed a mechanism for the spontaneous formation of pair-density waves, where the underlying superconductivity arises from pairing with a $\mathbb{Z}_2$ spin liquid. Screening of the spin liquid by the conduction sea then gives rise to triplet pair-density waves. While our model represents a specific case, the underlying concept extends to other forms of unconventional superconductivity believed to result from spin fractionalization, such as the RVB theory of high-$T_c$ superconductivity.

 
Pair-density wave (PDW) order has indeed been observed in STM experiments in the heavy fermion superconductor UTe$_2$ \cite{gu_detection_2023}, where it has been identified as the parent order underlying secondary charge-density wave (CDW) order, through melting transitions \cite{aishwarya_melting_2024}. Furthermore, time-reversal symmetry breaking observed in chiral step-edge experiments \cite{jiao_chiral_2020} and the onset of re-entrant superconductivity at 40T \cite{aoki_unconventional_2022}, alongside the absence of a two-stage transition and the fully gapped spectrum revealed by transport measurements \cite{suetsugu_fully_2024,theuss_single-component_2024}, have fueled debate \cite{jiao_chiral_2020,ishizuka_insulator-metal_2019,xu_quasi-two-dimensional_2019,machida_theory_2020,bae_anomalous_2021,hayes_multicomponent_2021,machida_nonunitary_2021,nakamine_inhomogeneous_2021,thomas_spatially_2021,fujibayashi_superconducting_2022,girod_thermodynamic_2022,rosa_single_2022,shaffer_chiral_2022,wei_interplay_2022,hazra_pair_2024,hazra_triplet_2023,iguchi_microscopic_2023,ishihara_chiral_2023,matsumura_large_2023,chang_correlation-driven_2024,suetsugu_fully_2024,theuss_single-component_2024} over the nature of superconducting order of UTe$_2$. This controversy arises because there are no known two-dimensional representations of superconducting order in orthorhombic(Immm) UTe$_2$, available to reconcile these experimental findings.

While our work does not provide a microscopic mechanism for the superconductivity in UTe$_2$, it points to a novel way out of this dilemma. Superconductivity breaks the U(1) isospsin symmetry associated with charge conservation.  In principle, group theory allows for both singly connected representations of the order parameter,  exemplified by the charge 2e BCS pairing, and  {\sl double-group} representations of the order parameter, with half-integer isospin.  Majorana-mediated superconductivity is a first concrete example of this second class of broken-symmetry. We may think of the electron-Majorana spinor as analogous to the quark
in strong interaction physics, and using this analogy, 
new families  of superconducting phases may then be possible, generated by product group representations akin to the eight-fold way of mesons.  In such a zoo of enriched superconducting phases, some may be topological\cite{zhuang_topological_2024}.  Importantly, double groups introduce Kramers degeneracy, allowing for broken time-reversal in low-symmetry crystalline environments, providing an alternative path for the emergence of spontaneous time-reversal symmetry-breaking from a heavy fermi liquid \cite{panigrahi_breakdown_2024}.

	\begin{acknowledgments}
		\textit{Acknowledgments:}
			AP and PC would like to thank Tamaghna Hazra for fruitful discussions. This work was supported by Office of Basic Energy Sciences, Material
		Sciences and Engineering Division, U.S. Department of Energy (DOE)
		under Contracts No. DE-SC0012704 (AMT) and DE-FG02-99ER45790 (AP and PC ).
	\end{acknowledgments}
	
	\newpage

\bibliographystyle{apsrev4-2}
\bibliography{PDWinCPT_X.bib}

\begin{widetext}
	\begin{center}
		\textbf{\large Supplemental Material for ``Microscopic theory of pair density waves in spin-orbit coupled Kondo lattice"}
	\end{center}
\end{widetext}
This supplementary material provides details of calculations on the
CPT model \cite{coleman_solvable_2022}, including the reduction from a four-band to a single band
model and the calculation of the finite momentum pair susceptibility.  The analytical tractability of this model, formulated on the hyperoctagon lattice, stems from nesting between the electron-hole and Majorana Fermi surfaces.

The Hamiltonian for the Kondo lattice (CPT) model consists of three components

\begin{equation} H_{CPT}=H_c+H_{YL}+H_K,
	\label{CPTHamAP} 
\end{equation}
where $H_c$ describes the nearest neighbor hopping of the conduction electrons, $H_{YL}$ describes the Yao-Lee spin-spin interaction and $H_K$ which couples the conduction sea to the Yao-Lee spin liquid via Kondo interaction,
\begin{align}
	H_{c}=&-t\sum_{\braket{i,j}} ( c\dg_{i\si}c_{j\si}+{\rm H.c.})-\mu\sum_{i}c\dg_{i\si}c_{i\si},\nonumber
	\\H_{YL}=&K/2\sum_{\braket{i,j}}\lambda^{\al_{ij}}_i\lambda^{\al_{ij}}_j
	(\vec{S}_i\cdot\vec{S}_j),\label{YLHAP}
	\\H_{K}=&J\sum_{i}(c\dg_{i}\vec{\si}c_{i})\cdot \vec{S}_{i}.
	\label{CPTCompAP} \nonumber
\end{align} 
	
	These three components of the Hamiltonian are as follows:
	\vskip 0.1 in
	
	\noindent {\sl \bf 1. Yao-Lee spin liquid:} The Yao-Lee term involves a Kitaev interaction between the orbitals $\lambda_j$ between nearest neighbors decorated with Heisenberg interaction between the spins $\vec{S}_j$. The result of the Kitaev orbital frustration is Majorana fractionalization of the orbitals $\vec{\lambda}_j=-i\vec{b}_j\times\vec{b}_j$ and the spins into $\vec{S}_j=-\frac{i}{2}\vec{\chi}_j\times\vec{\chi}_j$. 
	The physical Hilbert space is constrained by ${\cal D} = 8i b_1 b_2 b_3 \chi_1 \chi_2 \chi_3 = 1$, which commutes with the Hamiltonian.
	Consequently, the fractionalized representation of the Yao-Lee Hamiltonian is obtained by noting $\si^a_j\lambda^{\al}_j=2i \chi^a_j b^{\alpha}_j$ in the physical Hilbert space to be,
	\begin{equation}
		H_{YL}=K\sum_{\braket{i,j}}\hat{u}_{ij}(i\vec{\chi}_i\cdot \vec{\chi}_{j}).
	\end{equation}
	Where, similar to Kitaev spin liquid $\hat{u}_{ij}=i b^{\al_{ij}}_ib^{\al_{ij}}_j$ become static
	$\mathbb{Z}_2$ gauge fields, (i.e. $[H_{YL},\hat{u}_{ij}]=0$). 
	
	Noting that Yao-Lee spin liquid undergoes an Ising transition where below $T_c$\cite{hermanns_quantum_2014,coleman_solvable_2022,panigrahi_breakdown_2024}, leading to confinement of visons and deconfinement of Majoranas $\vec{\chi}_j$, we make the gauge choice $\hat{u}_{ij}=1$ \cite{hermanns_quantum_2014} leading to a translationally invariant Yao-Lee Hamiltonian. Transforming to
	a momentum basis, 
	\begin{equation} \vec{\chi}_{{\bf
				k},\alpha }=\frac{1}{\sqrt{N}}\sum_{j}\vec{\chi}_{j\alpha} e^{-i{\bf
				k}\cdot {\bf R}_j},
		\label{MajoranaFTAP} \end{equation} 
	where ${\bf R}_{j}$ is the position of the unit-cell in the BCC
	lattice, $N$ is the number of primitive unit cells in the lattice and $\alpha \in [1,4]$  is the site index within each unit cell. The Yao-Lee Hamiltonian is given by,
	\begin{equation}
		H_{YL}=K 
		\sum_{{\bf k}\in \cube }
		{\vec{\chi}}^{\dagger}_{\bf{k},\al}
		h({\bf k})_{\alpha,\beta}{\vec{\chi}}_{\bf{k}\beta},
		\label{YLHamAP}
	\end{equation}
	where (\ref{YLHamAP}) described a four-band Hamiltonian for three species of Majorana $\vec{\chi}_j$ with a global $SO(3)$ rotational symmetry, where 
	\begin{equation}
		\small\label{hkAP}
		h(\bf{k})=\begin{pmatrix}
			0 && i && i e^{-i \bf{k}\cdot \bf{a}_2} && i e^{-i \bf{k}\cdot \bf{a}_1} 
			\\-i && 0 && -i && i e^{-i \bf{k}\cdot \bf{a}_3}
			\\-i e^{i \bf{k}\cdot \bf{a}_3} && i && 0 && -i
			\\-i e^{i \bf{k}\cdot \bf{a}_1} && -i e^{i \bf{k}\cdot \bf{a}_2} && i && 0
		\end{pmatrix}.
	\end{equation}
	Moreover, the momentum spans $\cube\equiv 1/2 BZ$ the half-Brillouin zone. The Yao-Lee spin liquid can be described at low energies by an effective single-band model obtained by projecting to the lowest energy band. The dispersion for the said band is obtained by throwing away higher order terms in the characteristic equation for $h({\bf k})$, 
	\begin{equation}
		\tilde{\epsilon}^4-6\tilde{\epsilon}^2 -8\tilde{\epsilon} (s_{x}s_{y}s_{z})
		+ [ 4 (c_{x}^{2}+c_{y}^{2} + c_{z}^{2})-3]=0,
	\end{equation} 
	where, $c_i=\cos(k_i/2)$, $s_i=\sin(k_i/2)$ and $\tilde{\epsilon}$s are the eigenvalues of $h({\bf k})$.  
	
	The projected band describing the low energy physics of Yao-Lee Hamiltonian then takes the form, 
	\begin{equation} \label{ProjectedBand2AP}
		\tilde{\epsilon}({\bf k})=\frac{1}{2 s_{x}s_{y}s_{z}} \left[(c^2_x +c^2_y +c^2_z)-\frac{3}{4}\right],
	\end{equation}
	using which, the projected Yao-Lee Hamiltonian is expressed using $\epsilon_{\chi}({\bf k})=K\tilde{\epsilon}({\bf k})$ as follows,
	\begin{equation}\label{MajoranaHamAP}
		H_{YL}=\sum_{{\bf k}\in \cube} \epsilon_{\chi}({\bf k})  \vec{\chi}^{\dg}_{{\bf k}} \cdot \vec{\chi}_{{\bf k}}.
	\end{equation}
	\begin{figure}[h]
		\includegraphics[width=1.0\linewidth]{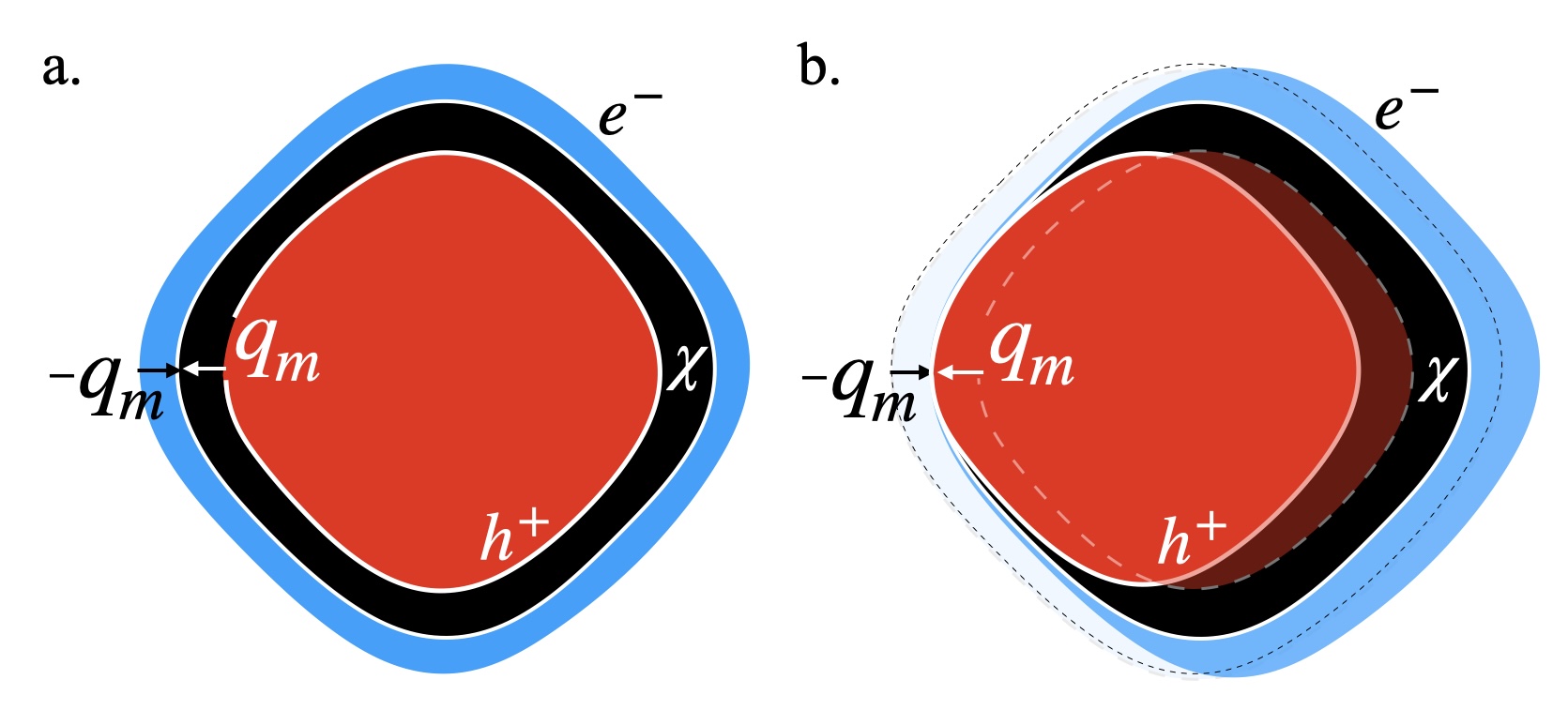}
		\caption{Illustrates the impact of electron doping on the conduction sea within a hyperoctagon lattice. (a)The conduction sea consists of  electron(blue) and hole Fermi surfaces (red), which expand and contract, respectively, as electron doping shifts the system away from half-filling. By contrast, the Majorana Fermi surface (black) is unaffected by doping.   (b) Doping stabilizes the development  of finite-momentum configurations of the electron-Majorana spinor order by shifting  the electron and hole Fermi surfaces shift by $\pm\vec{q}_m$ to nest with the Majorana Fermi surface.
		}
		\label{fig:Fermi_Surface}
	\end{figure}
	\vskip 0.1in
	
	\noindent {\sl \bf 2. Conduction electrons:} Similar to the
	Yao-Lee spin liquid, the conduction sea is also described
	by a four-band Hamiltonian, with an electron and a hole
	pocket (see \ref{fig:Fermi_Surface} ). To illustrate the perfect nesting between the conduction sea and the Majorana spinons, one carries out a non-singular gauge transformation $(c_1,c_2,c_3,c_4)_{\vec{R}}\rightarrow e^{i{\bf (\pi,\pi,\pi)\cdot R}} (c_1, i c_2,c_3,- i c_4)_{\vec{R}} $ which shifts the Fermi-surface to ${\bf Q}=(\pi,\pi,\pi)$ i.e., the $P$ point\cite{coleman_solvable_2022}. Following said gauge transformation, a four-band Hamiltonian describes the conduction sea,  
	\begin{equation}
		H_c=\sum_{{\bf k}\in {\bf BZ}}c^{\dagger}_{{\bf k},\si\al}[-t\  h({\bf k})-\mu \mathbb{I}]_{\al\beta}c_{{\bf k},\si\beta},
		\label{CondHamAP}
	\end{equation}
	where $h({\bf k})$ is given in equation (\ref{hkAP}).
	
	The low energy physics of the conduction band can be described by a projected single band $\epsilon_c({\bf k})=-t \tilde{\epsilon}({\bf k})$, with $\tilde{\epsilon}({\bf k})$ taking the form (\ref{ProjectedBand2AP}). Further, the conduction sea can be restricted to $1/2 BZ$ ($\cube$), by rewriting the Hamiltonian in terms of the Balian-Werthamer spinor  $\psi_{\bf k}=(c_{{\bf k}}, -i \sigma^2 c^{\dg}_{-{\bf k}})^T$ as,
	\begin{equation} \label{BLHcondAP}
		H_{c}=\sum_{{\bf k}\in \cube}\psi^{\dagger}_{{\bf k}}(\epsilon_{c}({\bf k})\mathbb{I}-\mu  \tau_3)\psi_{{\bf k}}.
	\end{equation}
	Here, coupling the chemical potential $\mu$ with $\tau_3$ indicates
	the presence of an electron and a hole Fermi surface in the conduction
	sea.
	\vskip 0.2in
	
	\noindent {\sl \bf 3. Kondo interaction:} The Majorana fractionalization of spins due to Yao-Lee interaction allows for an analytic mean-field treatment of Kondo interaction without resorting to Gutzwiller projection. In the Majorana representation the spins are expressed as $\vec{S}_j=-\frac{i}{2}\vec{\chi}_j\times\vec{\chi}_j$, which allows their rewriting as 
	\begin{equation}\label{KondoInteractionAP}\small
		H_{K}=
		J\sum_{j}(c\dg_{j}\vec{\sigma}c_j)\cdot (-\frac{i}{2}
		\vec{\chi}_j\times \vec{\chi}_j)\equiv - \frac{J}{2}\sum_{l}c\dg_{j}
		(\vec{\chi }\cdot \vec{ \sigma })^{2}c_{j}.
	\end{equation}
	The factorized form of the Kondo interaction (\ref{KondoInteractionAP}) allows for Hubbard Stratenovich transformation of the Kondo interaction in terms of a spinor order $V_j=-J\langle \vec{\chi}_a\cdot \vec{\si} c_j\rangle=(V_{j\up}, V_{j\dw})^{T}$,  
	\begin{equation}
		H_{K}=\sum_{j} \bigl[c\dg_{j} (\vec{\si}\cdot \vec{\chi}_j)V_j+{\rm H.c}\bigr] +\frac{2 V\dg_j V_j}{J}.
		\label{MeanFieldHamiltonianAP}
	\end{equation}
	Giving us the mean-field interaction term in equation (\ref{MeanFieldHamiltonianAP}).
	
	A more compact version of the mean-field Kondo interaction is obtained by switching to the Balian Werthamer spinor representation for conduction electrons $\psi_{k}=(c_{\vec{k}}, -i \sigma^2 c^{\dg}_{-\vec{k}})^T$ and the spinor order $\mathcal{V}_j= (V_j,-i\sigma^2 V^*_j)^T$ parameter,
	\begin{equation}\label{MFTHamAP}\small 
		H_{int}[j]=\frac{1}{2}\left(\psi^{\dg}_j (\vec{\si}\cdot \vec{\chi}_j) \mathcal{V}_j+\mathcal{V}^{\dg}_{j}(\vec{\si}\cdot\vec{\chi}_j)\psi_j \right)+\frac{\mathcal{V}^{\dg}_j \mathcal{V}_J}{J}.  
	\end{equation}

At half-filling, due to nesting between the electron, hole, and Majorana Fermi surfaces, the model exhibits a logarithmic divergence in the electron-Majorana susceptibility. This leads to the condensation of the uniform spinor order $V=\langle\vec{\chi_j}\cdot\vec{\sigma} c_j\rangle$ for infinitesimal Kondo coupling, akin to a Peierls instability.  
\vskip 0.2in
\noindent {\sl \bf Calculation of pair susceptibility at finite wavevector
	$\vec q$}
Away from half-filling, the nesting is disrupted by the presence of a non-zero chemical potential, which causes the electron Fermi surface to expand and the hole Fermi surface to contract, while the Majorana Fermi surface remains unaffected. The chemical potential thus acts as a cut-off for the logarithmic instability associated with superconductivity, stabilizing a finite momentum order parameter.

\begin{figure}[h]
	\centering
	\includegraphics[width=1\linewidth]{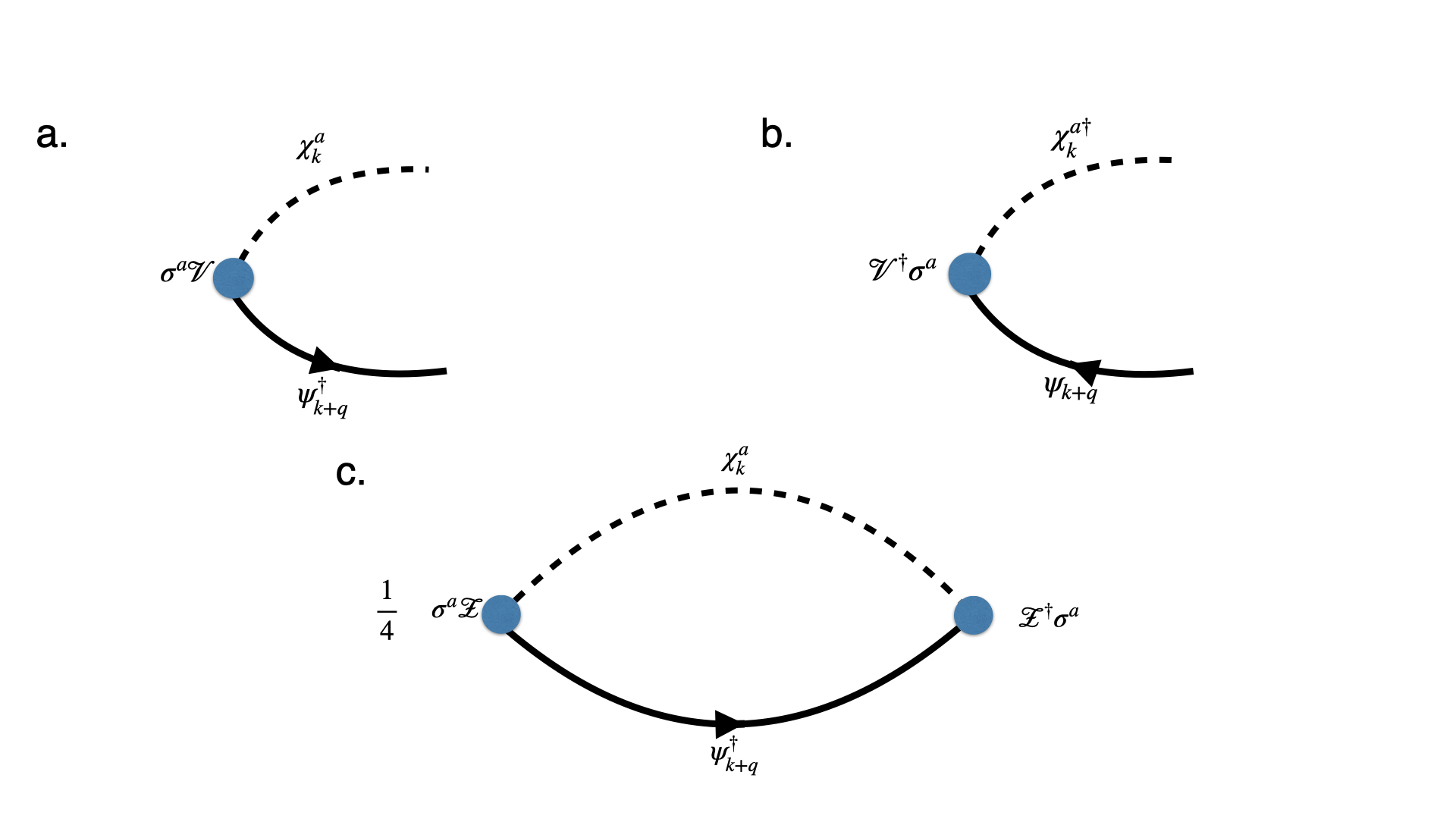}
	\caption{(a,b) Feynman diagrams for the second-order contribution to the free energy from the  Kondo hybridization between the conduction electron spinor $\psi^{\dagger}_{k+q}$ and the vector Majorana fields $\vec{\chi}_{k}$ with a finite momentum $q$ exchange. (c) Feynman diagram for the susceptibility bubble $\chi_{v}(q)$ related to electron-Majorana spinor ordering.} 
	\label{fig:SusceptibilityFD}
\end{figure}

This mechanism bears a strong analogy to singlet superconductors under an external magnetic field, where Zeeman splitting of the spin-polarized Fermi surface leads to the formation of Fulde-Ferrell-Larkin-Ovchinnikov (FFLO) \cite{fulde_superconductivity_1964,larkin_nonuniform_1964} pair-density waves. In both cases, pair-density waves emerge when the superconducting order parameter develops finite momentum.

The formation of such incommensurate pair density waves can be understood by examining the electron-Majorana susceptibility, which peaks at finite momentum. This susceptibility profile reveals that the ground-state energy of the system is minimized when the superconducting order parameter acquires finite momentum, thereby stabilizing the pair density wave phase.

Using a similar approach to FFLO, we show that the spinor order $V_j=\langle\vec{\chi_j}\cdot\vec{\si} c_j\rangle$ gains finite momentum in the analytically tractable Kondo lattice model \cite{coleman_solvable_2022} as one moves away from half-filling,  

\begin{equation}\label{unipodalAP}
	V_{j\si}=V e^{-i \vec{q}\cdot \vec{R}_j}z_{\si}.
\end{equation}
Here, $z_{\si}$ is the spinor orientation, and $V$ is the magnitude of the uniform order parameter with a spatial modulation with momentum $\vec{q}$.

Consequently, the mean-field Kondo interaction in Balian-Werthamer representation gains a finite momentum exchange,
\begin{equation} \label{MFTHam2AP}\small
	H_{int}=\sum_{j}\frac{1}{2}\left(\psi^{\dg}_j (\vec{\si}\cdot \vec{\chi}_j) e^{-i (\vec{q}\cdot \vec{R}_j){\tau_3}}\mathcal{V}+h.c.\right)+\frac{\mathcal{V}^{\dg} \mathcal{V}}{J}.  
\end{equation}

Here, $\tau_3$ in the exponent signifies that the holes gain opposite momentum to the electrons under Fourier transformation. 

Upon carrying out a Fourier transformation, the Kondo interaction in the momentum basis becomes,
\begin{equation}\small
	H_{int}=\sum_{\vec{k}\in \cube}\left(\psi^{\dg}_{\vec{k}+\vec{q}}(\vec{\si}\cdot \vec{\chi}_{\vec{k}})\mathcal{V}+h.c.\right)+\mathcal{N}\frac{\mathcal{V}^{\dg}\mathcal{V}}{J}.
\end{equation}
This compact notation expresses the exchange of $\vec{\chi}_k$ and $\psi^{\dg}_{\vec{k}+\vec{q}}$ Balian-Werthamer spinor which is diagrammatically depicted in Fig. \ref{fig:SusceptibilityFD}.         

In order to show that the finite momentum order parameter is preferred over the zero-momentum order parameter, one computes the susceptibility for spinor ordering. Where the finite momentum spinor order is preferred, the susceptibility is maximum at the said momentum $\vec{q}_m$.

Thus, to demonstrate the energetic favorability of the finite momentum order parameter, we begin by computing the static susceptibility (Fig. \ref{fig:SusceptibilityFD}) for electron-majorana condensation, 
\begin{widetext}
	\begin{equation}\label{Susceptibility1AP}\small
		\frac{\partial^2 F}{\partial V^2}=\chi_{v}(\vec{q})=-2T\sum_{i\omega_n,k\in \cube}\Tr\left[G_{\psi}(\vec{k}+\vec{q},i\omega_n)\sigma^a \mathcal{Z}\mathcal G_{\chi^a}(\vec{k},-i \omega_n)\mathcal{Z}^{\dagger}\sigma^a\right],
	\end{equation}
\end{widetext}
Where, $\mathcal{Z}=\frac{2\mathcal{V}}{V}$ is the orientation of the BW order parameter, $G_{\psi}(\vec{k},i\omega_n)=[(i\omega_n-\epsilon_c )\mathbb{I}_{[4]}+\mu\tau^3_{[4]}]^{-1}$ is the Green's function for the Balian-Werthamer spinor at Matsubara frequency $\omega_n$ and momentum $\vec{k}$, and $\mathbb{I}_{[4]}$ is $4$ dimensional Identity matrix, and $\tau^3_{[4]}=\mathbb{I}_{[2]}\otimes \tau^3 $ is the isospin matrix for Balian Werthamer representation, with $\vec{\tau}$ being Pauli matrices. And, $G_{\chi^a}(\vec{k},i\omega_n)=[i\omega_n-\epsilon_{\chi}(\vec{k})]^{-1}$ is the Green's function for $\chi^a$ Majorana with a Matsubara frequency $\omega_n$ and momentum $\vec{k}$.

Due to $SO(3)$ rotational symmetry of $\chi^a$ Majoranas, the Green's function $G_{\chi^a}$ being species independent, acts as an $I_{[3]}$ identity matrix, allowing us to rewrite the susceptibility as, 
\begin{widetext}
	\begin{equation}\label{SusceptibilitySimplifiedAP}\small
		\chi_{v}(\vec{q})=-2T\sum_{i\omega_n,k\in \cube}\Tr\left[\frac{1}{(i\omega_n-\epsilon_c(\vec{k}+\vec{q})) \mathbb{I}+\mu \tau^3}\frac{1}{i\omega_n+\epsilon_{\chi}(\vec{k})}\sigma^a \mathcal{Z}\mathcal{Z}^{\dagger}\sigma^a\right].
	\end{equation}
\end{widetext}
Further simplification is made by noting that $\mathcal{P}=\frac{1}{2}\mathcal{Z}\mathcal{Z}^{\dg}=\frac{1}{4}(1+d_{ab}\si^a\otimes\tau^b)$ and $\mathbb{I}_{[4]}-\mathcal{P}=\si^a \mathcal{Z}\mathcal{Z}^{\dg}\si^a=\frac{1}{4}(3-d_{ab}\si^a\otimes\tau^b)$ are projection operators\cite{coleman_odd-frequency_1994,coleman_solvable_2022,tsvelik_order_2022}. Moreover, $\si^a\mathcal{Z}\mathcal{Z}^{\dg}\si^a$ projects out a Majorana component $c^0$ of the conduction sea which remains gapless upon condensation of the fractionalized spinor order.
\begin{widetext}
	Additionally, introducing the electron-Majorana $\chi^{p}_{v}(\vec{q})$ susceptibility and hole-Majorana susceptibility $\chi^{h}_{v}(\vec{q})$,
	\begin{align}
		\chi^{p}_v (\vec{q})=&-2T\sum_{i\omega_n,k\in \cube}\frac{1}{(i\omega_n-\epsilon_c(\vec{k}+\vec{q})-\mu)(i\omega_n+\epsilon_{\chi}(\vec{k}))},
		\\ \text{and}, \quad \chi^{h}_v (\vec{q})=&-2T\sum_{i\omega_n,k\in \cube}\frac{1}{(i\omega_n-\epsilon_c(\vec{k}+\vec{q})+\mu)(i\omega_n+\epsilon_{\chi}(\vec{k}))}.
	\end{align}
	We obtain the following expression for the susceptibility to form fractionalized order:
	\begin{align}\label{Susceptibility2AP}
		\chi_{v}(\vec{q})&=\Tr\left[\left(\frac{\chi^{p}_v(\vec{q})+\chi^{h}_v(\vec{q})}{2}\mathbb{I}+\frac{\chi^{p}_v(\vec{q})-\chi^{h}_v(\vec{q})}{2}\tau^3\right)\frac{1}{4}(3-d_{ab}\si^a\otimes\tau^b)\right]=\frac{3}{2}[\chi^{p}_{v}(\vec{q})+\chi^{h}_{v}(\vec{q})]
	\end{align}
\end{widetext}
Thus, the task of calculating the susceptibility of the fractionalized order reduces to calculating the sum of electron-Majorana susceptibility $\chi^{p}_{v}(\vec{q})$ and hole-Majorana susceptibility $\chi^{h}_{v}(\vec{q})$. These susceptibilities are calculated by first carrying out the Matsubara summation over $i\omega_n$ to obtain 
\begin{equation}
	\chi^{p}_v (\vec{q})=-\frac{1}{4}\sum_{k\in \cube}\frac{\tanh(\epsilon_{\chi}(\vec{k}))/2T+ \tanh(\epsilon_c(\vec{k}+\vec{q})+\mu)/2T}{\epsilon_{\chi}(\vec{k})+\epsilon_c(\vec{k}+\vec{q})+\mu},
	\label{momentumsusceptibility}
\end{equation}
\text{and}, 
\begin{equation}
	\chi^{h}_v (\vec{q})=-\frac{1}{2}\sum_{k\in \cube}\frac{f(\epsilon_{\chi}(\vec{k}))-f(\epsilon_c(\vec{k}+\vec{q})-\mu)}{\epsilon_{\chi}(\vec{k})-\epsilon_c(\vec{k}+\vec{q})+\mu}.
\end{equation}
Where $f(\epsilon)$ is the Fermi-Dirac function. 

We rewrite the momentum integral along the constant energy contours to compute the spinor-susceptibility. This is done by noting that the density of states of the electrons and Majoranas around the Fermi surface remains constant. Further, since the Fermi-Dirac function $f(\epsilon)=\Theta(-\epsilon)$ becomes a Heavyside-Theta function at zero temperature i.e. $T=0$, the electron-Majorana susceptibility becomes,
\begin{equation}\small
	\chi^{p}_v (\vec{q})=-\frac{\rho(0)}{2}\int d\Omega\int d\epsilon \frac{\Theta(-K\epsilon)+\Theta(-t\epsilon+\mu)-1}{(K+t)\epsilon+\vec{v}_F(\theta,\phi)\cdot \vec{q}-\mu}.
\end{equation}
Where $\rho=\frac{2\sqrt{3}}{\pi^2 t}$ \cite{coleman_solvable_2022} is the density of states and $D$ is the bandwidth of the conduction sea.
Here, we have carried out a Taylor expansion $\epsilon_{c}(\vec{k}+\vec{q})=\epsilon_{c}(\vec{k})+\vec{q}\cdot\vec{v}_F(\theta,\phi)$ to obtain the above expression for electron-Majorana susceptibility, with the Fermi-velocity $\vec{v}_F=\vec{\nabla}_{\vec{k}}\epsilon_c(\vec{k})\vert_{\epsilon_F}$ being the gradient of the dispersion at the Fermi surface. 

Accounting for the theta functions, the energy integral takes the form:
\begin{equation}\small
	\chi^{p}_v (\vec{q})=-\frac{\rho(0)}{2}\int d\Omega\left[\int_{-D}^{0}-\int_{\mu/t}^{D} \right] \frac{d\epsilon}{(K+t)\epsilon+\vec{v}_F(\theta,\phi) \cdot\vec{q}-\mu},
\end{equation}
Where the energy integral is carried out analytically to obtain the following expression for $\chi^{p}_{v}(\vec{q})$,
\begin{widetext}
	\begin{equation}\small
		\chi^{p}_v (\vec{q})=-\frac{\rho(0)}{2(K+t)}\int d\Omega\ln\left(\left\vert(K+t)\epsilon+\vec{v}_F(\theta,\phi)\cdot \vec{q}-\mu\right\vert\right)\left[\vert_{-D}^{0}-\vert_{\mu/t}^{D} \right] . 
	\end{equation}
	Taking the limits for the energetic integral followed by a simplification of the above equation yields the following expression for the susceptibility,
	\begin{equation}\label{HyperoctagonSusceptibiltyAP}
		\small \chi_v (\vec{q})=\frac{6\rho(0)}{(K+t)}\ln\left(\frac{(K+t) D}{\mu}\right)-\frac{3\rho(0)}{2(K+t)}\int \frac{d\Omega}{4\pi}\left[\ln\left(\left\vert 1-\frac{\vec{v}_F(\theta,\phi)\cdot \vec{q}}{\mu}\right\vert\right)+\ln\left(\left\vert \frac{K}{t}+ \frac{\vec{v}_F(\theta,\phi)\cdot \vec{q}}{\mu}\right\vert \right)\right].
	\end{equation}
\end{widetext}
Where, $\rho(0)$ is the density of states for $\tilde{\epsilon}(\vec{k})$, and the spinor order susceptibility $\chi_{v}(\vec{q})=\frac{3}{2}(\chi^{p}_{v}(\vec{q}+\chi^{h}_{v}(\vec{q}))$, as given in equation (\ref{Susceptibility2AP}). This susceptibility maximizes for $\vec{q}\in \{q_m(\pm 1, 0,0), q_m( 0,\pm 1,0),q_m( 0,0,\pm 1)\}$ with $q_m=0.77\pi \mu$. These orders are degenerate, and the breaking of these degeneracy by higher order terms in the Ginzburg Landau theory leads to amplitude modulation of the spinor-order, which subsequently leads to amplitude modulation of triplet pairing.

To assess the stability of the pair density wave state at finite temperature, we note that when the susceptibility is maximized at a finite momentum $\vec{q}_m$, it must be convex around $\vec{q} = 0$. Expanding the susceptibility as $\chi^p_v(T,\mu,\vec{q}) \approx a + b q^2 + \mathcal{O}(q^4)$, the condition $b > 0$ characterizes the onset of the PDW phase. To compute this expansion, one considers the finite-temperature susceptibility $\chi^p_v(T,\mu,\vec{q})$ given in equation (\ref{momentumsusceptibility}) and rewrites it in the form of an energy integral.
\begin{widetext}
	\begin{equation}
		\chi_{v}^p(T,\mu,\vec{q})= \frac{\rho(0)}{4}\int d\varepsilon \int d\Omega \frac{\left(\tanh\left( \frac{\varepsilon}{2T}+\frac{\mu^*}{4T}\right)+ \tanh\left(\frac{\varepsilon}{2T}-\frac{\mu^*}{4T} \right)\right)}{2\varepsilon+\vec{v}_F(\Omega)\cdot\vec{q}}
	\end{equation}
	Where we take the limit $t=K=1$ and $\mu^*=t\mu$. In order to further simplify the susceptibility we carry out the substitution $x=\frac{\varepsilon}{2T}$, which yields, $2T dx=d\varepsilon$. We then expand the denominator in orders of $\vec{q}$  and carry out the angular integration to obtain:
	\begin{equation}
		\chi_{v}^p(T,\mu,\vec{q})= \frac{\rho^*(0)}{4}\int dx  \left(\tanh\left( x+\frac{\mu^*}{4T}\right)+ \tanh\left(x-\frac{\mu^*}{4T} \right)\right)\left(\frac{1}{x}+\frac{\kappa}{x^3}\left(\frac{\vec{v}_F\vec{q}}{4T}\right)^2\right)+O(q^4)
	\end{equation}
	We have used the angular averages $\langle \vec{v} \cdot \vec{q} \rangle_{\Omega} = 0$ and $\langle (\vec{v} \cdot \vec{q})^2 \rangle_{\Omega} = \kappa q^2 > 0$ for the linear and quadratic terms, respectively.
	
	Consequently, the quadratic coefficient of $q^2$ in the susceptibility expansion is given by:
	\begin{equation}
		b\left(\frac{\mu}{4T}\right)=\frac{\kappa\rho^*(0)}{4}\left(\frac{v_F}{4T}\right)^2 \int dx \frac{\tanh\left( x+\frac{\mu^*}{4T}\right)+ \tanh\left(x-\frac{\mu^*}{4T} \right)}{x^3}
		\label{SMChi}
	\end{equation}

\end{widetext}

Here, $\kappa$ encodes the geometry of the Fermi surface. Finally, since the prefactor of the integral is positive, the overall sign is determined by the integral itself, which, on numerical evaluation, changes sign at $\mu^*/4T = 0.9552$.

\begin{figure}
	\includegraphics[width=1.0\linewidth]{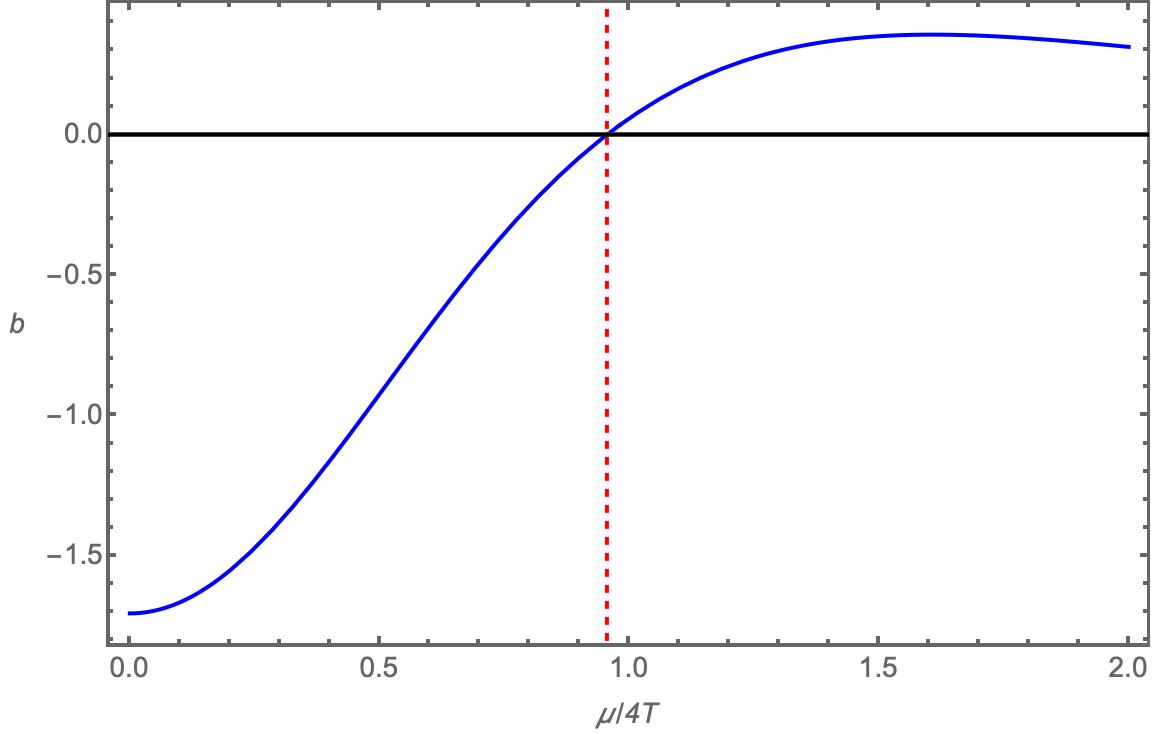}
	\caption{Illustrates the plot between $b/b_0$ (i.e. the integral sans the prefactor) as a function of $\mu/4T$. The sign change at $\mu/4T=0.9552$ signifies the point below which the pair density wave phase becomes unstable to thermal fluctutation and the system transitions from a normal state to a uniform triplet superconductor instead of inhomogeneous triplet pair density wave state. 
	}
	\label{fig:BvsQPlot}
\end{figure}

Since the chemical potential cuts off the logarithmic instability in the PDW phase, the transition occurs at a finite Kondo coupling $ J_c $. For small chemical potentials, the required Kondo coupling remains sufficiently low due to Stoner's criterion, $ \chi_{v}(\vec{q}) = \frac{1}{J_c}$. This ensures that the model remains analytically tractable over a range of chemical potentials $ \mu $. Consequently, this model is one of the rare examples where the spontaneous formation of pair-density waves can be reliably established without the need for an external field beyond one-dimension.
\vskip 0.2in
\noindent {\sl \bf Local electronic self-energy, non-collinear spinors and the intertwined order}
\vskip 0.1in
{
	
	The non-collinear nature of the spinor order $\mathcal{V}_i$ gives rise to a rich variety of exotic odd-frequency modulations, including spin density waves, charge density waves, and both singlet and triplet pair density wave (PDW) modulations. These emerge as direct consequences of the underlying spinor order modulation and provides a unified microscopic framework for understanding intertwined phases. To explore the character of these intertwined orders, we begin by considering the local electronic self-energy:
	\begin{equation}
		\Sigma(x,\omega)=\sum_a\si^a\mathcal{V}(x)\mathcal{V}^{\dg}(x)\si^a \mathcal{G}_\chi(\omega),
	\end{equation}
	where $\mathcal{G}_\chi(\omega)$ is the local propagator for the Majorana Fermions in the Yao-Lee Spin liquid. We note that since $\mathcal{G}_\chi(\omega) = - \mathcal{G}_\chi(-\omega)$, the scattering self-energy is odd-in frequency. 
	
	As our susceptibility analysis indicates, finite chemical potential $\mu$ favors spinor-order modes at finite momentum in the ground state. The interference between such modes leads to modulations in a variety of electronic observables. To illustrate this, we analyze a bimodal scenario for the spinor order $\mathcal{Z}_i= {\cal V}_i/V$:
	\begin{equation}
		\mathcal V(x) = V \mathcal Z(x)= V(Z_1e^{i \vec q \cdot \vec x}+ Z_2e^{-i qx})
	\end{equation}
	The local self-energy is then given by 
	\begin{equation}
		\Sigma(x,\omega) = V^2 G_{\chi}(\omega) A(x)
	\end{equation}
	This results in the modulation of the self-energy via $ A(x)=\sigma^a\mathcal Z(x)\mathcal Z^{\dg}(x)\sigma^a $, (where the repeated roman index $a=1,2,3$ implies $\sum_{a=1,3}$).
	We now focus on the modulated part $\mathcal A$ of ${A}$:
	\begin{equation}\label{amx}
		\mathcal A(x)= \sigma^a\mathcal{Z}_1 \mathcal{Z}^{\dg}_2\sigma^a e^{2i \vec q\cdot \vec x}+ \sigma^a\mathcal{Z}_2 \mathcal{Z}^{\dg}_1\sigma^ae^{- 2i \vec q \cdot \vec x}.
	\end{equation}
	We note that $\sigma^a\mathcal Z_1 \mathcal Z_2^{\dg}\sigma^a$, is spanned by $\si^{\al}\tau^{\beta}$, where $\sigma^{\alpha}=(\sigma^0,\sigma^1,\sigma^2,\sigma^3)$ and $\tau^{\alpha}=(\tau^0,\tau^1,\tau^2,\tau^3)$ are the four component Pauli matrices
	for spin and isospin, respectively. We then write
	$
	\sigma ^a\mathcal Z_1 \mathcal Z_2^{\dg}\sigma^a=d_{\al\beta}\si^{\al}\tau^{\beta}$, $\sigma^a\mathcal Z_2 \mathcal Z_1^{\dg}\sigma^a=\tilde d_{\al\beta}\si^{\al}\tau^{\beta}$, 
	where
	\begin{eqnarray}\label{dees}
		d_{\al\beta}&=&\mathcal Z^{\dg}_2 \sigma^a(\si^{\al}\tau^{\beta})\sigma^a\mathcal Z_1=
		h_{\alpha}\mathcal Z^{\dg}_2 \si^{\al}\tau^{\beta}\mathcal Z_1, \cr  
		\tilde d_{\al\beta}&=& \mathcal Z^{\dg}_1 \sigma^a (\si^{\al}\tau^{\beta})\sigma^a\mathcal Z_2=
		h_{\alpha} \mathcal Z^{\dg}_1 \si^{\al}\tau^{\beta}\mathcal Z_2.
	\end{eqnarray}
	Here the signature $h = (3, -1, -1, -1) $ results from the anticommutations between the sigma matrices in $\sigma^a\sigma^\alpha \sigma^a = h_{\alpha} \sigma^{\alpha}$ (there is no implied summation on $\alpha$).
	It follows that $\mathcal A(x)= \mathcal A_{\alpha\beta}(x)\sigma^{\alpha}\tau^{\beta}$, where
	\begin{equation}\label{amxz}
		\mathcal A_{\alpha\beta}(x) =  d_{\alpha\beta}e^{2 i \vec q \cdot \vec x}+\tilde d_{\alpha\beta}e^{-2 i \vec q \cdot \vec x}
	\end{equation}
	
	Now the $\tilde d_{\alpha\beta}$ satisfy the  relations
	\begin{equation}\label{crucial}
		\tilde d_{\alpha\beta } =  d_{\alpha\beta}^*=g_{\alpha}g_{\beta}d_{\alpha\beta},
	\end{equation}
	where the signature $g=(1,-1,-1,-1)$. 
	The first relation follows directly from \eqref{dees}. 
	To prove the second equality, take the transpose of the expression for $\tilde d_{\alpha\beta}$ in \eqref{dees},
	$\tilde d_{\alpha\beta } =  h_{\alpha }\mathcal Z^T_2 (\si^{\alpha}\tau^{\beta})^T \mathcal Z^*_1$,
	where $\mathcal Z^* = (\mathcal Z\dg)^T$.  Now under transposition, 
	$
	(\sigma^{\alpha}\tau^{\beta})^T= g_{\alpha}g_{\beta} C (\sigma^{\alpha} \tau^{\beta}) C 
	$, where $C = \sigma_2\tau_2$, so it follows that 
	\begin{eqnarray}
		\tilde d_{\alpha\beta } &=&  h_{\alpha} g_{\alpha}g_{\beta}\mathcal Z^T_2 C(\si^{\alpha}\tau^{\beta})C \mathcal Z^*_1\cr
		&=& h_{\alpha}g_{\alpha}g_{\beta}\mathcal Z\dg_2 (\si^{\alpha}\tau^{\beta}) \mathcal Z_1=  g_{\alpha}g_{\beta}d_{\alpha\beta}.
	\end{eqnarray}
	where we have used the   transposition
	law ${\mathcal C} \mathcal Z^*= -\mathcal Z $,  $\mathcal Z^T{\mathcal C} = -\mathcal Z\dg $, from which \eqref{crucial} follows.  Substituting \eqref{crucial} into \eqref{amxz} we obtain
	\begin{equation}
		\mathcal A_{\alpha\beta}(x) = \left\{ \begin{array}{cl}\ \ {2} {\rm Re} (d_{\alpha\beta})\cos 2\vec q \cdot \vec x,  & g_{\alpha}g_{\beta} = +1\cr
			-{2} {\rm Im}(d_{\alpha\beta}) \sin 2\vec q \cdot \vec x, & g_{\alpha}g_{\beta} = -1
		\end{array}\right.
	\end{equation}
	From the structure of the amplitudes, we conclude that the  modulated component of the self-energy takes the form $\Sigma(x,\omega) = 
	G_{\chi}(\omega) {\Sigma_m}(x)$, where 
	\begin{widetext}
		
		\begin{equation}
			{\Sigma^{m}_{\alpha\beta}(x)}=
			\begin{array}{c|c|ccc}
				\alpha \backslash \beta & 0 & 1 & 2 & 3 \\
				\hline
				0 & 
				\Sigma_{00} \cos (2 \vec q \cdot \vec x)& 
				\mathrm{Re}(\Delta_s) \sin (2 \vec q \cdot \vec x)& 
				\mathrm{Im}(\Delta_s) \sin (2 \vec q \cdot \vec x) &
				\mu \sin (2 \vec q \cdot \vec x) \\[6pt]
				\hline
				1 & 
				m_x \sin (2 \vec q \cdot \vec x)&
				d_{1x} \cos (2 \vec q \cdot \vec x) &
				d_{2x}\cos (2 \vec q \cdot \vec x) &
				d_{3x}\cos (2 \vec q \cdot \vec x)
				\\[6pt]
				
				2 & 
				m_y \sin (2 \vec q \cdot \vec x) &
				d_{1y}\cos (2 \vec q \cdot \vec x) &
				d_{2y}\cos (2 \vec q \cdot \vec x) &
				d_{3y}\cos (2 \vec q \cdot \vec x)\\[6pt]
				
				3 & 
				m_z \sin (2 \vec q \cdot \vec x) &
				d_{1z}\cos (2 \vec q \cdot \vec x) &
				d_{2z}\cos (2 \vec q \cdot \vec x) &
				d_{3z}\cos (2 \vec q \cdot \vec x) \\
			\end{array}
		\end{equation}
	\end{widetext}
	We see that the modulation in the triplet pairing components $\vec d_i$ occurs with $\cos(2 \vec q \cdot  \vec x)$ dependence, while the  modulation of the charge density term $\mu$, spin density wave term $\vec{m}$, and the singlet pairing modulation ($\Delta_s$) occurs with a sine wave $\sin(2 \vec q \cdot \vec x)$. The appearance of a $\pi/2 $ phase shift between the primary  triplet pairing and the secondary spin, charge and singlet pair modulations is a key feature of the theory.

	\begin{acknowledgments}
		\textit{Acknowledgments:}
		This work was supported by Office of Basic Energy Sciences, Material
		Sciences and Engineering Division, U.S. Department of Energy (DOE)
		under Contracts No. DE-SC0012704 (AMT) and DE-FG02-99ER45790
		(PC and AP).
		
	\end{acknowledgments}
	
	\bibliographystyle{apsrev4-2}
	\bibliography{PDWinCPT_X}

    \end{document}